# PERIGALACTIC DISTANCES OF GLOBULAR CLUSTERS

SIDNEY VAN DEN BERGH

Dominion Astrophysical Observatory

5071 West Saanich Road

Victoria, British Columbia

V8X 4M6, Canada

Electronic Mail:  vandenbergh@dao.nrc.ca

**ABSTRACT**

Luminosities and tidal radii of globular clusters have been used to estimate the perigalactic distances P of well-observed Galactic globulars that do not have collapsed cores. It is found that the cluster metallicity [Fe/H] correlates somewhat more strongly with P than it does with the present Galactocentric distances R of globular clusters. It is also found that both P and R correlate more strongly with the cluster half-light radii $r_h$ than they do with [Fe/H]. This suggests that cluster radii are mainly determined by global parameters of the (proto)Galaxy, whereas [Fe/H] may have been more strongly affected by local evolutionary effects. The strong correlation $\rho$ (P) = 0.70 ± 0.05 between the perigalactic distances and half-light radii of globular clusters is particularly striking. The half-light radii $r_h$ of globular clusters in the outer halo of the Galaxy are found to correlate with R, and even more strongly with P. It is not yet clear how this result is to be reconciled with the Searle-Zinn scenario for the formation of the outer halo of the Galaxy. The tidal radii of clusters with collapsed cores appear to have been systematically over-estimated, resulting in perigalactic distances that are too large. Halo globulars with P ≳ 15 kpc are (with the exception of NGC 2419) typically an order of magnitude less luminous than the clusters with P < 15 kpc that are associated with the main body of the Galaxy. The population of nuclear globular clusters with P < 1 kpc appears to be deficient in objects fainter than $M_V \approx -7$. Clusters with well-determined ages having P > 15 kpc have ages in the range 9 - 16 Gyr, whereas globulars with P < 15 kpc all appear to have ages that are restricted to the narrower range $13 \leq T (Gyr) \leq 16$.

## 1. INTRODUCTION

Globular clusters are the oldest sub-units of galaxies. They therefore provide valuable information on the earliest phases of galactic evolution. In the present paper an attempt will be made to study the relationships between the perigalactic distances P and other characteristics of globular clusters. It would be particularly interesting to know if cluster metallicity correlates more strongly with P than it does with the present Galactocentric distance R. Does the correlation between [Fe/H] and P extend into the halo? What is the relationship between the ages of globular clusters and their perigalactic distances? Answers to these and other questions might provide interesting constraints on scenarios for the early evolution of the Milky Way system.

## 2. PERIGALACTIC DISTANCES

King (1962) has shown that globular clusters may exhibit well-defined tidal radii $r_t$. According to Freeman & Norris (1981), such tidal radii are given approximately by

$$r_t \propto R \, (M_c/M_G)^{1/3} \qquad (1)$$

in which $M_c$ and $M_G$ are the cluster mass and the Galactic mass interior to R, respectively. For a galaxy with a constant rotation curve $M_G \propto R$, so that (e.g. Goodman 1993)

$$r_t \propto R^{2/3} \, M_c^{1/3} \,. \qquad (2)$$

For a more detailed discussion of the relation between tidal radii, perigalactic distances and Galactic mass models, the reader is referred to Innanen, Harris & Webbink (1983). Both the measurement and interpretation of tidal radii are, however, complicated by recent observations which suggest that some globular clusters appear to exhibit outer distortions and faint tidal tails (Grillmair et al. 1995).

Clusters on elongated orbits will be most severely truncated near perigalacticon, i.e. when R = P. Eqn. (2) therefore provides a way of estimating the perigalactic distances of Galactic globular clusters. However, Oh & Lin (1992) caution that diffusion might significantly modify the structure of the outer regions of some globular clusters in which the two-body relaxation time is moderately short. In such clusters the limiting radius may, in fact, be comparable to the tidal radius at apogalacticon. Furthermore, Oh (1994) finds that significant numbers of stars on retrograde orbits may exist beyond the cluster tidal radius. Additional complications are due to observational uncertainty in $r_t$, which may result from small errors in the adopted level of background counts, and from the fact that possible differences between the mass-to-light ratios of individual globular clusters make the transformation from luminosity $M_V$ to total cluster mass $M_c$ uncertain.

Despite these potential problems, an attempt will be made to derive the perigalactic distances P from Eqn. (2), the log $r_c$ and c ≡ log ($r_t/r_c$) values compiled by Trager, Djorgovski & King (1993) [only clusters for which these parameters were of quality classes 1 and 2 were used], and the present Galactocentric distances R recently tabulated by Djorgovski (1993). The constant of proportionality in Eqn. (2) was, more-or-less arbitrarily, set in such a way that only 15% of the clusters that did not have collapsed cores have P > R. (Since P ≤ R for real cluster orbits, values of P > R must be mostly due to observational errors and/or variations in the *M*/L ratios of globular clusters.) If one assumes that only 10% of all clusters have P > R, then the calculated perigalactic distances are all systematically decreased by 20%. On the other hand, the assumption that 20% of all clusters have P > R results in an increase of 26% in the calculated perigalactic distances of all clusters. Note that the rank correlation coefficients used in §§ 7 and 8 will not be affected by scale errors in P.

Values of the perigalactic distance P (in kpc) calculated from Eqn. (2) are listed in Table 1. Values of P for clusters with P > R have been placed in parentheses. Values of P for clusters having collapsed cores (c ≥ 2.0), which are probably biased, have been marked with an asterisk.

Agreement between the perigalactic distances quoted in Table 1 and those calculated by Allen & Santillán (1993) from cluster proper motions and two Galactic mass models is poor. In seven out of eight cases, Allen & Santillán find larger values of P than those that are calculated from Eqn. (2). For all eight objects in common, the Allen & Santillán scale of P vales is, on average, 2.9 times larger than that adopted in the present study. If such a large correction were applied to the present P values, then most Galactic globular clusters would have perigalactic values larger than their present Galactocentric distances! The perigalactic distances for two clusters with measured proper motions calculated by Odenkirchen & Brosche (1992) are also somewhat larger than those given for the same objects in Table 1. For NGC 6218 (M 12), which is the only object in common to all three studies, P (A & S) = 4.3 kpc, P (O & B) = 2.1 kpc and P (vdB) = 1.4 kpc.

From new proper motion reductions Peterson, Rees & Cudworth (1995) find P $\approx$ 1 kpc for NGC 6121 (M4), which agrees well with the value P = 1.4 kpc given in Table 1.

A check on the adopted scale of P values is provided by those globular clusters which, from their radial velocity and position on the sky, are found to be in nearly circular orbits (von Hoerner 1955, van den Bergh 1993). Such clusters should have values of R/P $\approx$ 1. This is indeed found to be the case for NGC 5139, NGC 5904, NGC 6362 and NGC 6723, which are the only four globulars with nearly circular orbits listed in Table 1. For these objects it is found that $<\log R - \log P> = 0.17 \pm 0.09$. This confirms that the adopted scale for perigalactic distances cannot be greatly in error.

An additional check on the significance of the differences between values of P and R for individual globular clusters is provided by looking only at those 15 clusters in Table 1 for which van den Bergh (1993) finds plunging (i.e. highly elongated) orbits. For all 15 such clusters $<\log R - \log P> = 0.51$. Excluding NGC 6401, which has an exceptionally large value $R/P = 79$, one obtains $<\log R - \log P> = 0.41 \pm 0.07$. This is significantly larger than the value $<\log R - \log P> = 0.17 \pm 0.09$ found for those globular clusters to which van den Bergh (1993) assigns circular orbits. It is therefore concluded that statistical data on the orbital shapes derived from the computed perigalactic distances are entirely consistent with those derived from cluster radial velocities, in conjunction with distances and position on the sky, as derived by the method of Perek (1954) and von Hoerner (1955).

### 3. CLUSTERS WITH COLLAPSED CORES

The distribution of log (R/P) for normal clusters, and for clusters with collapsed cores $[c \equiv \log (r_t/r_c) \geq 2.0]$, is compared in Fig. 1. A Kolmogorov-Smirnov test shows that there is $< 0.1\%$ probability that the core-collapsed, and non core-collapsed cluster samples were drawn from the same parent population of log (R/P) values. A possible explanation for this difference is that the value $\log (r_t/r_c) = c = 2.5$ adopted for collapsed-core clusters by Trager et al (1993) is too large and should, in fact, be ≈ 2.0. As a result, the tidal radii calculated from log $r_t$ = log $r_c$ + c, and hence the values of the perigalactic distance P, are probably biased for clusters with collapsed cores. An additional complication (Lee & Goodman 1995) is introduced by the fact that post core-collapse clusters may have unusually high stellar evaporation rates. Because of these uncertainties, the P values calculated for compact clusters with c ≥ 2.0 have not been used in the subsequent discussion. A plot of log (R/P) versus c ≡ log ($r_t$/$r_c$) is shown in Fig. 2. This Figure illustrates the excess of clusters with P > R for objects with c ≥ 2.0.

## 4. METALLICITY VERSUS P AND R

Figure 3 shows a plot of [Fe/H], taken from a recent compilation by Djorgovski (1993), versus log P. Clusters with collapsed cores have not been plotted in the Figure because their perigalactic distances are believed to suffer from systematic bias. Also shown in Fig. 3 is the relation between [Fe/H] and log R for the same sample of clusters. Intercomparison of these two plots shows that the metallicity of globular clusters appears to be somewhat more strongly correlated with the perigalactic distance P than with the present Galactocentric distance R. For the data on [Fe/H] and P, one obtains a Spearman (1904) rank correlation coefficient ρ (P) = 0.47 ± 0.08. (This rank correlation coefficient is, of course, independent of any scale error in our adopted values of P.) For the same clusters, the data on [Fe/H] and R give a much lower value $\rho(R) = 038 \pm 0.09$. This shows that <u>the metallicity of Galactic globular clusters correlates marginally more strongly with the perigalactic distance P than it does with the present Galactocentric distance R</u>. This result strengthens and confirms a similar result previously obtained by Seitzer (1983), which is quoted in Freeman & Norris (1981). Clusters on elongated orbits spend more time near apogalacticon than they do near perigalacticon. A plot of [Fe/H] versus log R therefore provides information that is more closely associated with on the relation between metallicity and apogalactic distance. Taken at face value, Fig. 3 therefore suggests that <u>the metallicities of globular clusters are somewhat more closely tied to their perigalactic distances than they are to their distances at apogalacticon</u>. Could this indicate that globular clusters typically formed near perigalacticon (where gas densities were high), rather than near apogalacticon (where gas densities were lower)? It appears premature to try to answer this question at the present time.

Mean values of log P and log R in four metallicity bins are given in Table 2. The data in this Table show a monotone increase of $<\log P>$ with decreasing metallicity over the range [Fe/H] $\simeq$ -0.5 to [Fe/H] $\simeq$ -2.0. The dependence of $<\log R>$ on [Fe/H] appears to be less clear-cut than that of $<\log P>$ on [Fe/H].

Inspection of the data in Table 2 shows no obvious dependence of the difference $<\log R> - <\log P>$ on metallicity, i.e. there seems to be no obvious dependence of orbit shape on metallicity. A similar conclusion had previously been drawn by Innanen et al (1983) from their study of Galactic globular cluster orbits.

## 5. GLOBULAR CLUSTER AGES

Recently, Richer et al (1995) have given a compilation of information on 35 globular clusters that have accurate modern age determinations. In Fig. 4, these ages are plotted as a function of perigalactic distance P. Although presently available data are not numerous, they do appear to suggest that outer halo globular clusters with $P \gtrsim 15$ kpc have a wide range of ages. On the other hand, globular clusters with $P < 15$ kpc, which are associated with inner halo or with the main body of the (proto)Galaxy, seem to be restricted to ages in the relatively narrow range 13 - 16 Gyr. The difference between halo globulars and clusters associated with the main body of the Galaxy appears much less clear-cut in the plot of cluster age versus present Galactocentric distance R. It is tentatively concluded that <u>cluster ages correlate more clearly with perigalactic distances than they do with their present Galactocentric distances</u>.

## 6. LUMINOSITIES OF GLOBULAR CLUSTERS

Do the luminosities of Galactic globular clusters depend on their perigalactic distances? Figure 5 suggests that such a relationship does, in fact, exist. All halo clusters (except NGC 2419) that have P > 15 kpc are seen to be fainter than $M_V \approx$ -6. Typical halo clusters are seen to be an order of magnitude (2 - 3 mag) fainter (less massive?) than the average of globulars located at smaller perigalactic distances. Inspection of Fig. 5 also suggests, but does not prove, that the population of clusters with perigalactic distances P < 1 kpc is deficient in objects fainter than $M_V \approx$ -7. Possibly, such low-luminosity clusters were destroyed by bulge shocks (Aguilar, Hut & Ostriker 1988). Alternatively, such faint objects might have evaded discovery among the dense star and dust clouds that are seen in the direction of the Galactic center.

## 7. DIAMETERS OF GLOBULAR CLUSTERS

A number of recent investigations (e.g. van den Bergh, Morbey & Pazder 1991, van den Bergh 1994) have shown that the half-light radii $r_h$ of globular clusters grow with increasing galactocentric distance R. This raises the question whether $r_h$ might correlate even more strongly with the perigalactic distance P. These two correlations are intercompared in Fig. 6. Inspection of this Figure shows that $r_h$ does, in fact, appear to correlate more strongly with P than it does

- 13 -

with R. For the clusters listed in Table 1, it is found that the Spearman (1904) rank correlation coefficient $\rho$ (P) = 0.70 ± 0.05, which is strikingly larger than the value $\rho$ (R) = 0.55 ± 0.08. It is therefore concluded that <u>the correlation between the half-light radii of globular clusters with perigalactic distance is significantly stronger than that which had previously been found between half-light radii and the present Galactocentric distances of clusters</u>. It is also of interest to note that Table 3 shows P and R to correlate more strongly with $r_h$ than with [Fe/H]. This suggests that cluster radii are mainly determined by global Galactic parameters, whereas [Fe/H] may have been more strongly affected by local evolutionary effects.

In Fig. 6, globular clusters that are probably on retrograde orbits (van den Bergh 1993) are plotted as open circles. The Figure shows that clusters on retrograde orbits have below-average radii. Perhaps by chance the correlation (r = 0.89 ± 0.09) between log $r_h$ and log P appears to be particularly tight for clusters on retrograde orbits.

8.     **HALO GRADIENTS**

Many years ago, Searle (1978) pointed out that the metallicity of globular clusters in the range 10 < R (kpc) < 50 appeared to be independent of Galactocentric distance. This important, and at that time unexpected, result led Searle & Zinn (1978) to propose that the outer halo of the Galaxy had formed by prolonged infall of protogalactic fragments. The data collected in Table 4 confirm Searle's conclusion that metallicity and Galactocentric distance are not correlated (and might, in fact, even show marginal evidence for an anti-correlation) for the 13 globular clusters (that do not have collapsed cores) which are listed in Table 1 and have R > 20 kpc. A similar conclusion is seen to hold for the 13 globular clusters that have P > 20 kpc. However, a very different situation is found to prevail for the half-light radii of the outer halo globular clusters that are listed in Table 1. For these objects, a correlation at the ~ 3 $\sigma$ level of significance is found between R and $r_h$ for both (1) the 13 clusters with R > 20- kpc, and (2) the 13 clusters with P > 10 kpc.

Finally, an even stronger correlation is seen in Table 4 between the half-light radii of halo globular clusters and their perigalactic distances. For the 13 clusters with P > 10 kpc $\rho$ (P, $r_h$) = +0.84 ± 0.08 (m.e), while clusters with R > 20 kpc yield $\rho$ (P, $r_h$) = +0.83 ± 0.08. These results show that <u>the half-light radii of globular clusters exhibit a highly significant gradient in the outer Galactic halo</u>. Since this gradient is largely due to the absence of compact globulars at the largest distances, it seems implausible that this gradient was mainly produced by selective destruction of fragile clusters by disk shocks and Galactic tidal forces. It is not yet clear how the existence of strong cluster diameter gradients in the outer Galactic halo is to be reconciled with the Searle-Zinn paradigm.

## 9. CONCLUSIONS

Information on the perigalactic distances of Galactic globular clusters provides a number of tantalizing clues about the early evolutionary history of the Galaxy. The age distribution of globular clusters suggests that the main body (or inner halo) of the protogalaxy at $P < 15$ kpc only formed globular clusters during the period between 13 Gyr and 16 Gyr ago. On the other hand, some outer halo globulars with $P > 15$ kpc appear to have ages as low as ~ 10 Gyr. An additional difference between globulars in the Galactic core and halo is that clusters associated with the main body of the Galaxy have $<M_V> \approx -7.5$, compared to $<M_V> \approx -5.25$ for most outer halo clusters. Furthermore, the cluster metallicity [Fe/H] appears to be slightly more correlated with perigalactic distance P than it is with the present Galactocentric distance R. The interpretation of this result is not yet clear. It is also found that both P and R correlate more strongly with the half-light radii $r_h$ of clusters than they do with [Fe/H]. Finally, it is noted that $r_h$ correlates with P in the outer Galactic halo, even though [Fe/H] and P do not.

I am indebted to Kyle Cudworth, Jeremy Goodman and Scott Tremaine for providing me with important references, to David Duncan for drawing the Figures and to Janet Currie for typing the manuscript.

**TABLE 1 - PERIGALACTIC DISTANCES OF GLOBULAR CLUSTERS**

| Name   | log P   | Name   | log P   | Name   | log P    |
|--------|---------|--------|---------|--------|----------|
| N 104  | 0.43*   | N5694  | 0.58    | N6304  | 0.29     |
| N 228  | 0.63    | I4499  | 1.06    | N6316  | -0.10    |
| N 362  | 0.46    | N5824  | 1.18*   | N6325  | 0.24*    |
| N1261  | 0.44    | Pal. 5 | (1.72)  | N6341  | 0.40     |
| AM 1   | 1.60    | N5897  | 0.72    | N6333  | 0.04     |
|        |         |        |         |        |          |
| Eri.   | 1.52    | N5904  | 0.67    | N6342  | (0.82)*  |
| Ret.   | (1.78)  | N5946  | (0.82)* | N6356  | 0.32     |
| N1851  | 1.10*   | N5986  | 0.30    | N6355  | (0.34)*  |
| N1904  | 0.37    | Pal. 14| (1.88)  | N6352  | 0.26     |
| N2298  | 0.39    | N6093  | 0.38    | Ter. 2 | (0.96)*  |
|        |         |        |         |        |          |
| N2419  | 1.30    | N6121  | 0.14    | N6366  | 0.48     |
| N2808  | 0.27    | N6139  | 0.12    | N6362  | (0.72)   |
| E 3    | (0.96)  | N6171  | 0.52    | HP 1   | (0.09)*  |
| Pal. 3 | 1.64    | N6205  | 0.62    | Lil. 1 | (0.19)*  |
| N3201  | 0.70    | N6218  | 0.16    | N6380  | 0.28     |
|        |         |        |         |        |          |
| Pal. 4 | 1.50    | N6229  | 0.62    | N6388  | -0.24    |
| N4147  | 0.80    | N6235  | 0.28    | N6402  | (0.84)   |
| N4590  | (1.02)  | N6254  | 0.36    | N6401  | -1.64    |
| N4833  | 0.63    | N6256  | 0.46*   | N6397  | -0.18*   |
| N5024  | 1.09    | Pal. 15| 1.32    | Pal. 6 | 0.09     |
|        |         |        |         |        |          |
| N5053  | 1.05    | N6266  | -0.34   | N6426  | (1.16)   |
| N5139  | 0.40    | N6273  | 0.29    | Ter. 5 | (0.11)*  |
| N5272  | 0.98    | N6284  | (1.08)* | N6440  | -0.32    |
| N5286  | 0.08    | N6287  | (0.37)  | N6441  | -0.08    |
| N5634  | 0.84    | N6293  | (0.53)* | N6453  | (1.04)*  |

| Name   | log P    | Name   | log P    | Name    | log P   |
|--------|----------|--------|----------|---------|---------|
| UKS 1  | (0.86)*  | N6652  | 0.10     | N7006   | 0.99    |
| N6496  | -0.02    | N6656  | 0.14     | N7078   | 0.63*   |
| N6517  | -0.44    | Pal. 8 | 0.14     | N7089   | 0.74    |
| N6522  | (0.80)*  | N6681  | 0.26*    | N7099   | 0.56*   |
| N6535  | 0.34     | N6712  | -0.04    | Pal. 12 | (1.28)  |
| N6528  | (0.68)*  | N6715  | 0.31     | Pal. 13 | 1.22    |
| N6539  | 0.40     | N6717  | 0.34*    | N7492   | 1.22    |
| Djo. 3 | 0.61*    | N6723  | 0.26     | -       | -       |
| N6544  | -0.10*   | N6749  | -0.02    | -       | -       |
| N6541  | (0.71)*  | N6752  | (0.86)*  | -       | -       |
| N6553  | -0.43    | N6760  | 0.37     |         |         |
| N6558  | (0.36)*  | N6779  | 0.34     |         |         |
| I1276  | (0.84)   | Pal. 10| (1.24)   |         |         |
| N6569  | -0.02    | Arp 2  | (1.67)   |         |         |
| N6584  | 0.36     | N6809  | 0.25     |         |         |
| N6624  | (0.66)*  | Pal. 11| 0.70     |         |         |
| N6626  | 0.00     | N6838  | 0.12     |         |         |
| N6638  | 0.18     | N6864  | 0.44     |         |         |
| N6637  | 0.22     | N6934  | 0.57     |         |         |
| N6642  | (0.40)   | N6981  | 0.79     |         |         |

**TABLE 2 - METALLICITY VERSUS LOG R AND LOG P**

| [Fe/H] | $<\log R>$ | $<\log P>$ | $[R/P]^a$ | $n^b$ |
|---|---|---|---|---|
| $\leq -1.75$ | $1.09 \pm 0.09$ | $0.76 \pm 0.11$ | $0.33 \pm 0.14$ | 22 |
| -1.74 to -1.25 | $1.02 \pm 0.08$ | $0.64 \pm 0.09$ | $0.38 \pm 0.12$ | 40 |
| -1.24 to -0.75 | $0.58 \pm 0.09$ | $0.30 \pm 0.20$ | $0.28 \pm 0.22$ | 13 |
| $> -0.75$ | $0.64 \pm 0.07$ | $0.10 \pm 0.07$ | $0.54 \pm 0.10$ | 17 |

[a] $[R/P] \equiv <\log R> - <\log P>$

[b] Number of clusters

**TABLE 3 - SPEARMAN RANK CORRELATION COEFFICIENTS**

| Method | $\rho$ |
|---|---|
| $r_h$ and P | $0.70 \pm 0.05$ |
| $r_h$ and R | $0.55 \pm 0.08$ |
| [Fe/H] and P | $0.47 \pm 0.08$ |
| [Fe/H] and R | $0.38 \pm 0.09$ |

**TABLE 4 -  RANK CORRELATION COEFFICIENTS FOR CLUSTERS IN THE OUTER HALO**

| Parameters | $\rho$ |
|---|---|
| P > 10, R and [Fe/H] | $-0.38 \pm 0.24$ |
| P > 10, P and [Fe/H] | $-0.46 \pm 0.22$ |
| P > 10, R and $r_h$ | $+0.63 \pm 0.17$ |
| P > 10, P and $r_h$ | $+0.84 \pm 0.08$ |
| R > 20, R and [Fe/H] | $-0.11 \pm 0.27$ |
| R > 20, P and [Fe/H] | $+0.13 \pm 0.27$ |
| R > 20, R and $r_h$ | $+0.61 \pm 0.18$ |
| R > 20, P and $r_h$ | $+0.83 \pm 0.08$ |

## FIGURE LEGENDS

Fig. 1  Distribution of log (R/P) values for clusters with collapsed cores ($c \geq 2.0$), and for clusters that do not have collapsed cores. The fact that the majority of clusters with collapsed cores are calculated to have P > R suggests that the perigalactic distances of clusters with collapsed cores have been systematically over-estimated.

Fig. 2  Plot of log (R/P) versus $c \equiv \log (r_t/r_c)$ for individual Galactic globular clusters. The excess of objects with P > R at large values of c suggests that tidal radii and, hence log ($r_t/r_c$), have been systematically over-estimated for compact clusters with collapsed cores.

Fig. 3  Comparison between plots of metallicity [Fe/H] versus perigalactic distance P (top), and of [Fe/H] versus present Galactocentric distance R (bottom). The Figure suggests that metallicity may be somewhat more strongly correlated with P than it is with R.

Fig. 4  Perigalactic distance (top) versus age for clusters with accurately known ages. Clusters with collapsed cores have not been plotted. The Figure

suggests that globulars of all ages occur at P > 15 kpc. Only clusters with ages greater than 13 Gyr appear to occur among clusters with P < 15 kpc. The present Galactocentric distances R of clusters (bottom) may show the same effect, albeit less convincingly. Six clusters with collapsed cores are plotted with plus (+) symbols.

Fig. 5  Absolute magnitude $M_V$ versus perigalactic distance P for Galactic globular clusters. All halo clusters with P > 15 kpc (except NGC 2419) are seen to be ~ 2 - 3 mag fainter than the clusters with P < 15 kpc, which are associated with the main body of the (proto) Galaxy. Clusters for which P > R are shown as plus (+) symbols.

Fig. 6  Correlation between cluster half-light radii $r_h$ (in pc) and perigalactic distance P (in kpc) at top, and with present Galactocentric distance R at bottom. The Figure suggests that the correlation with perigalactic distance is marginally stronger than that with Galactocentric distance. Clusters on retrograde orbits (shown as open circles) appear to have below-average radii.

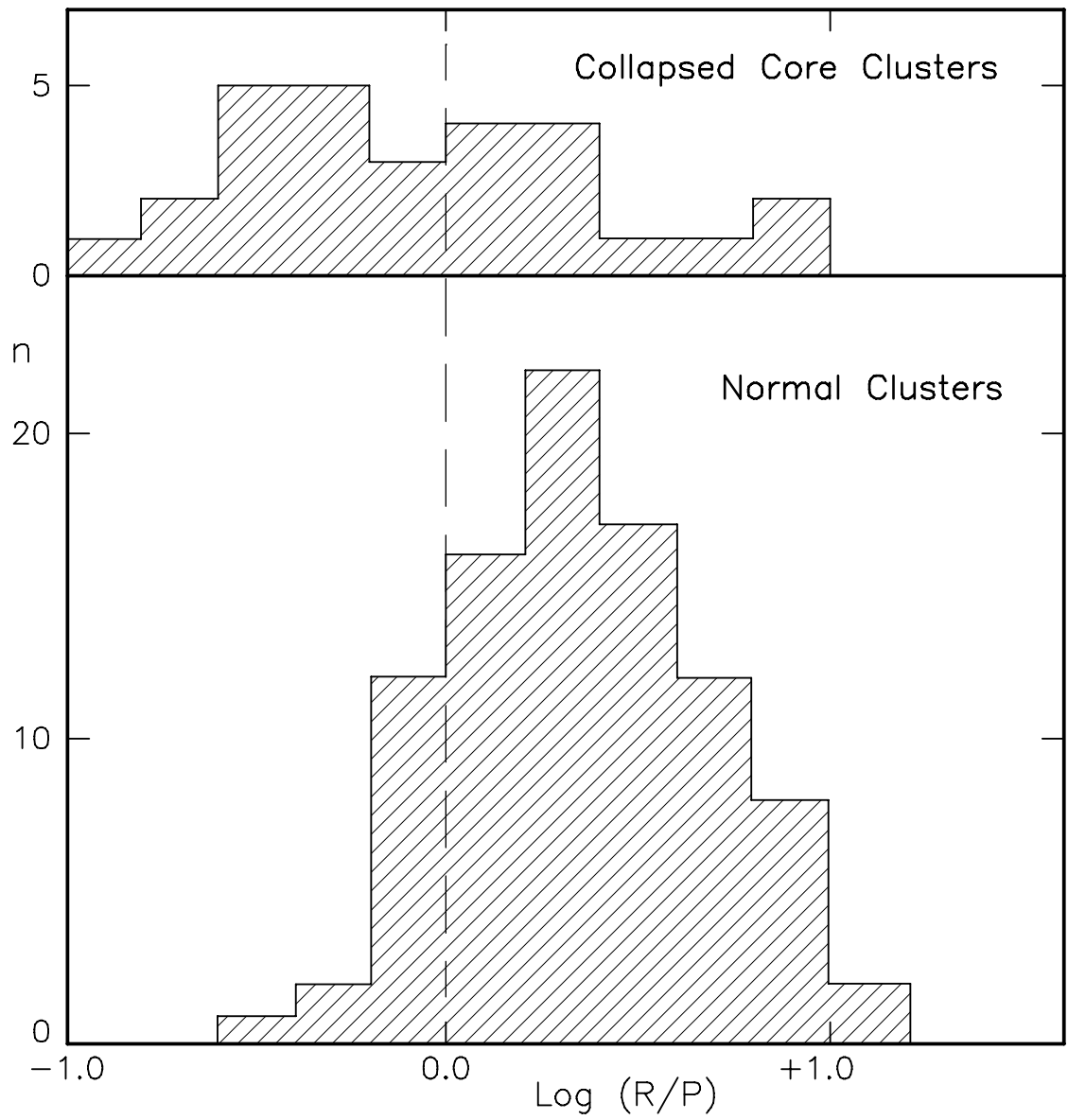

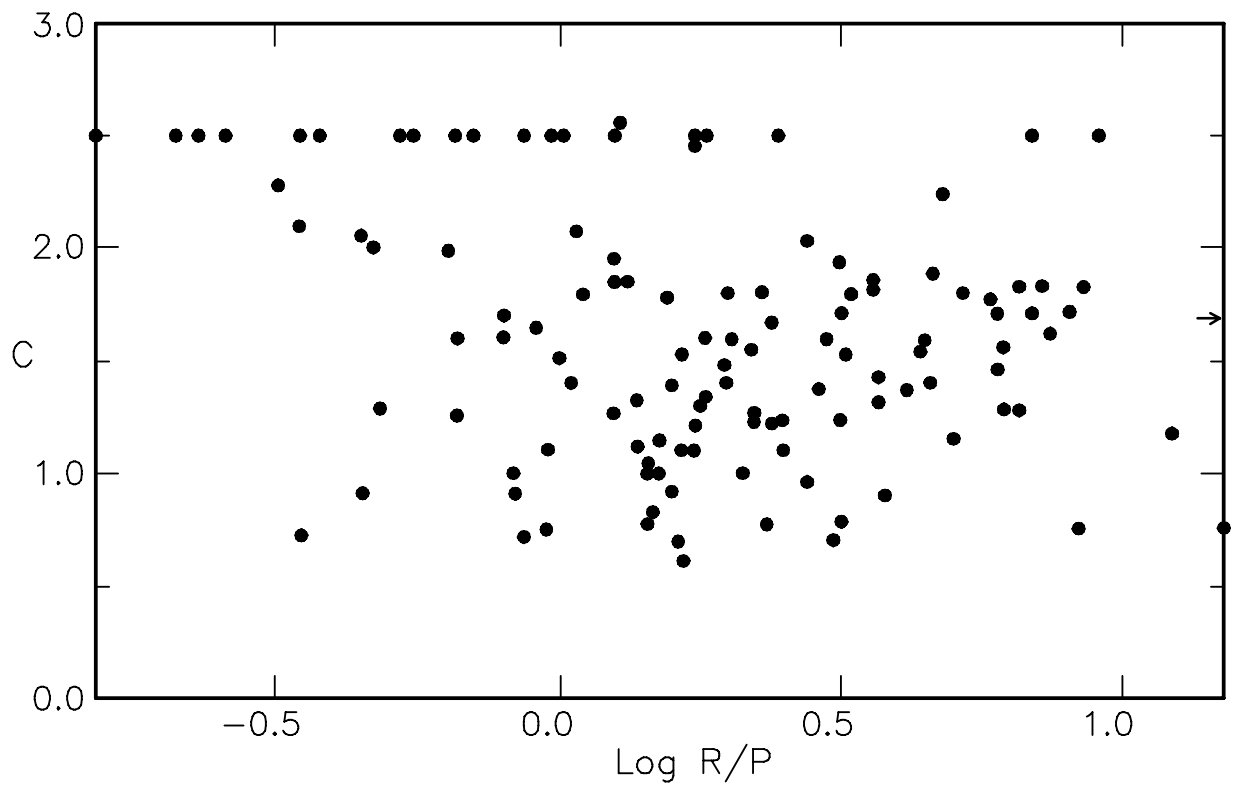

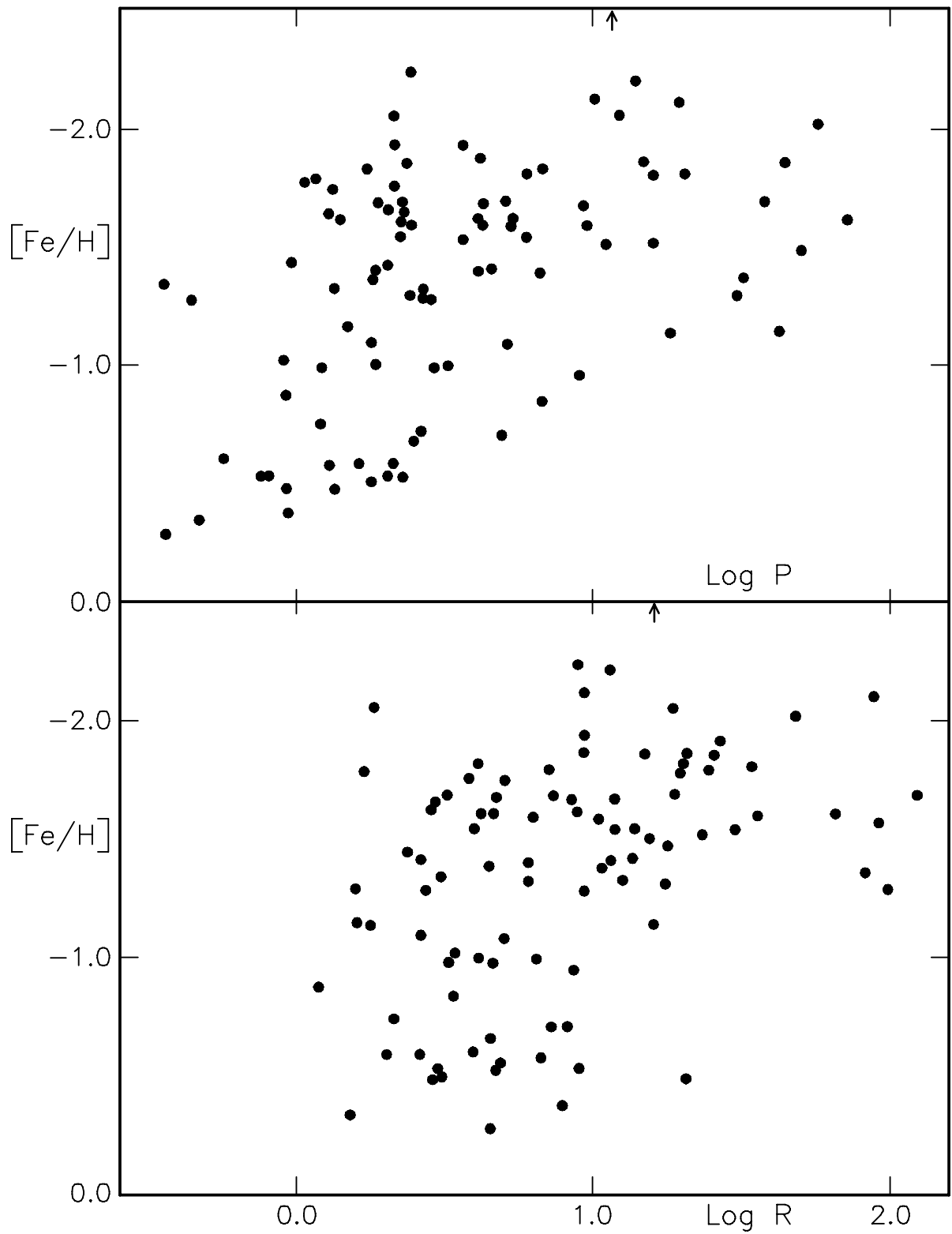

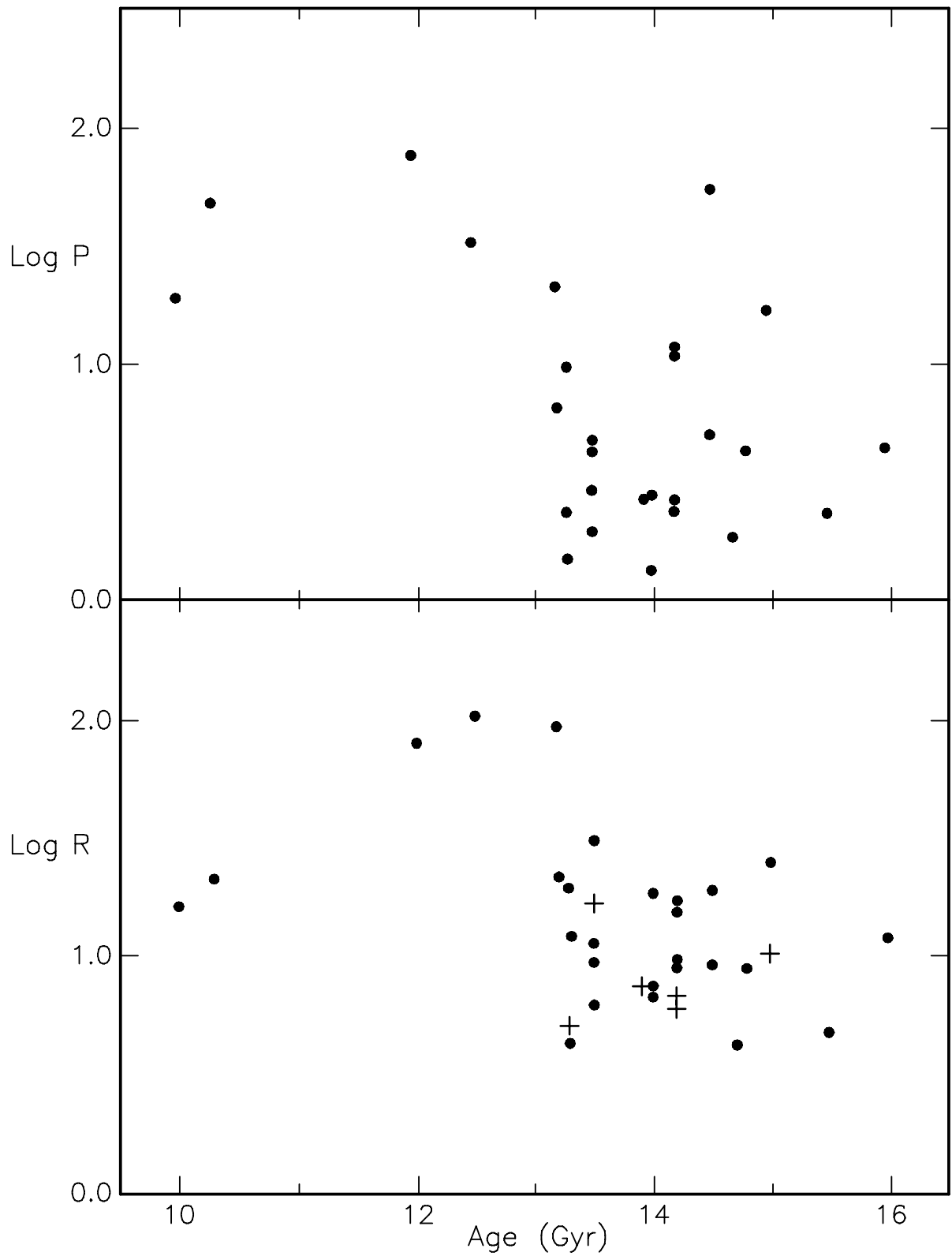

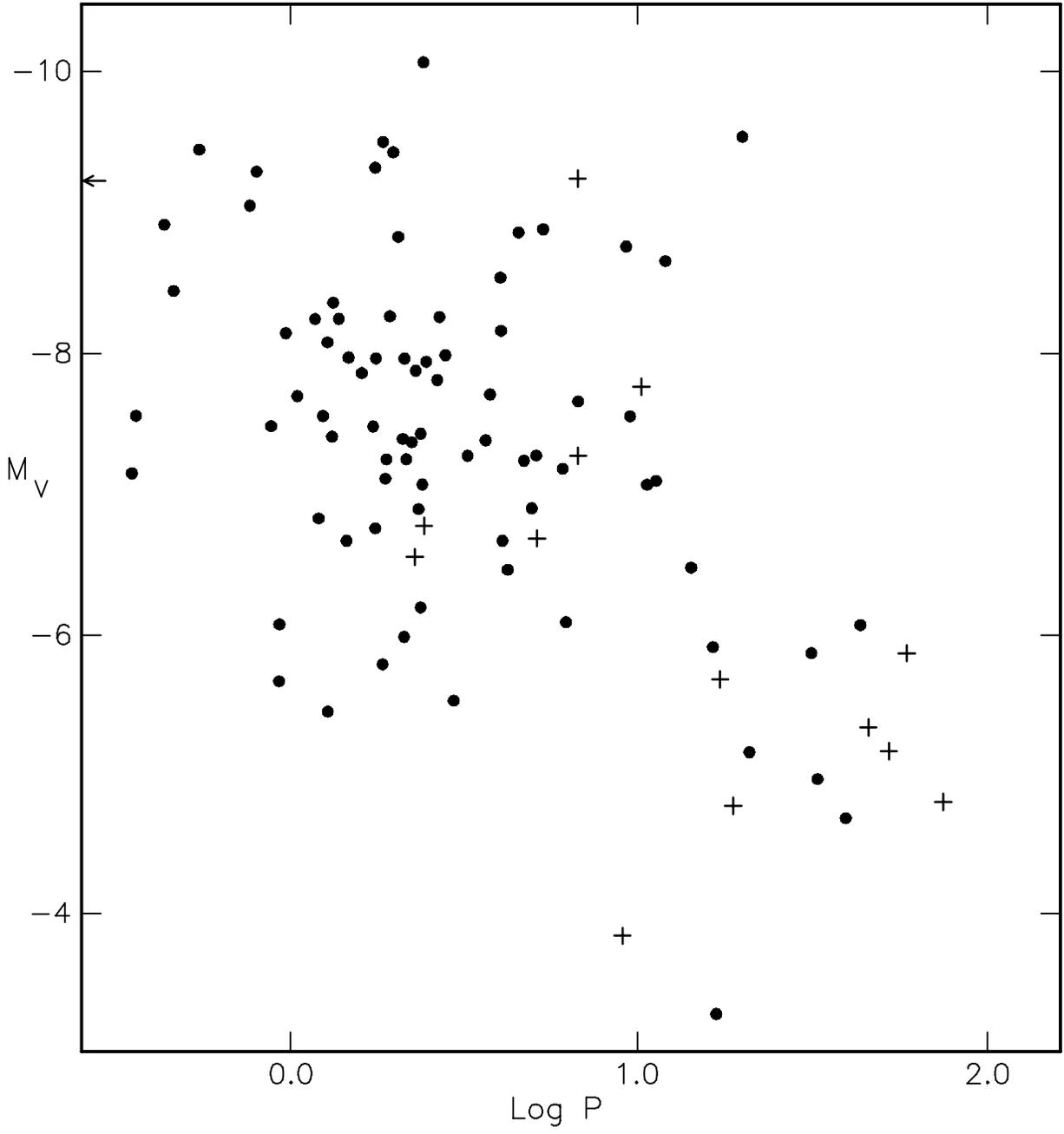

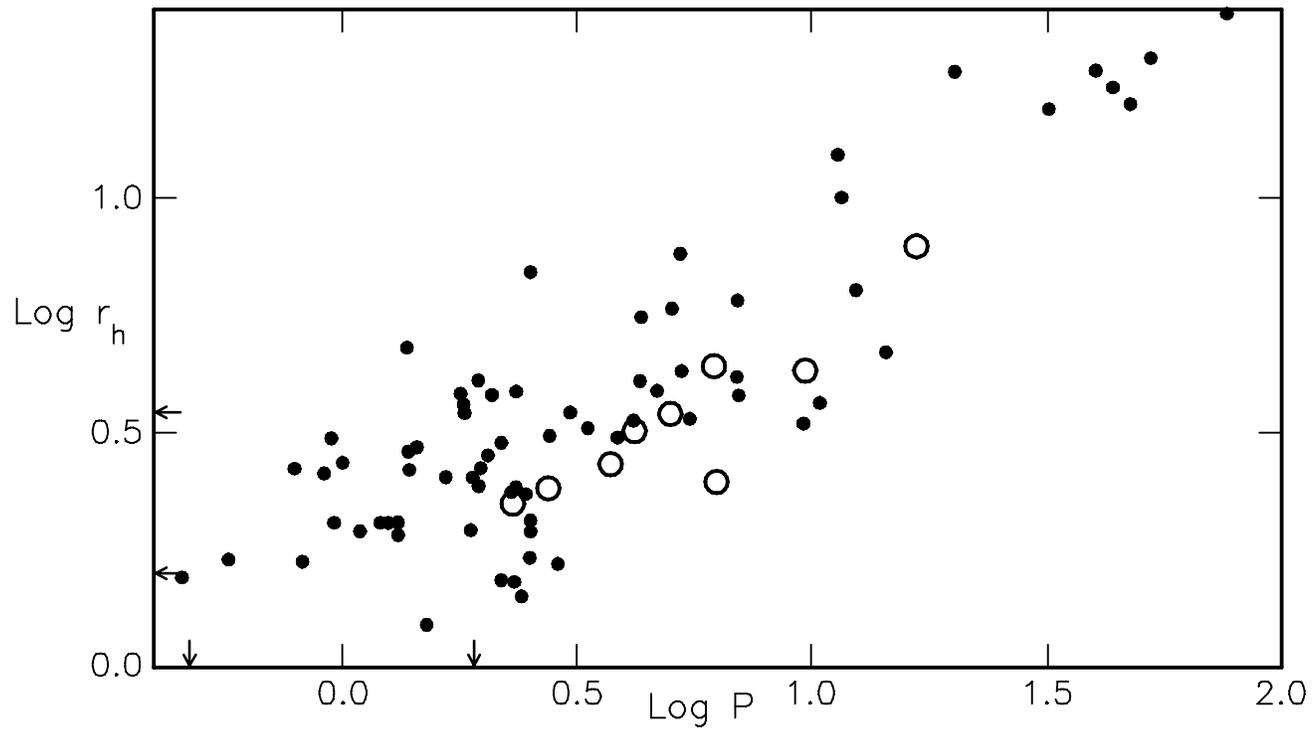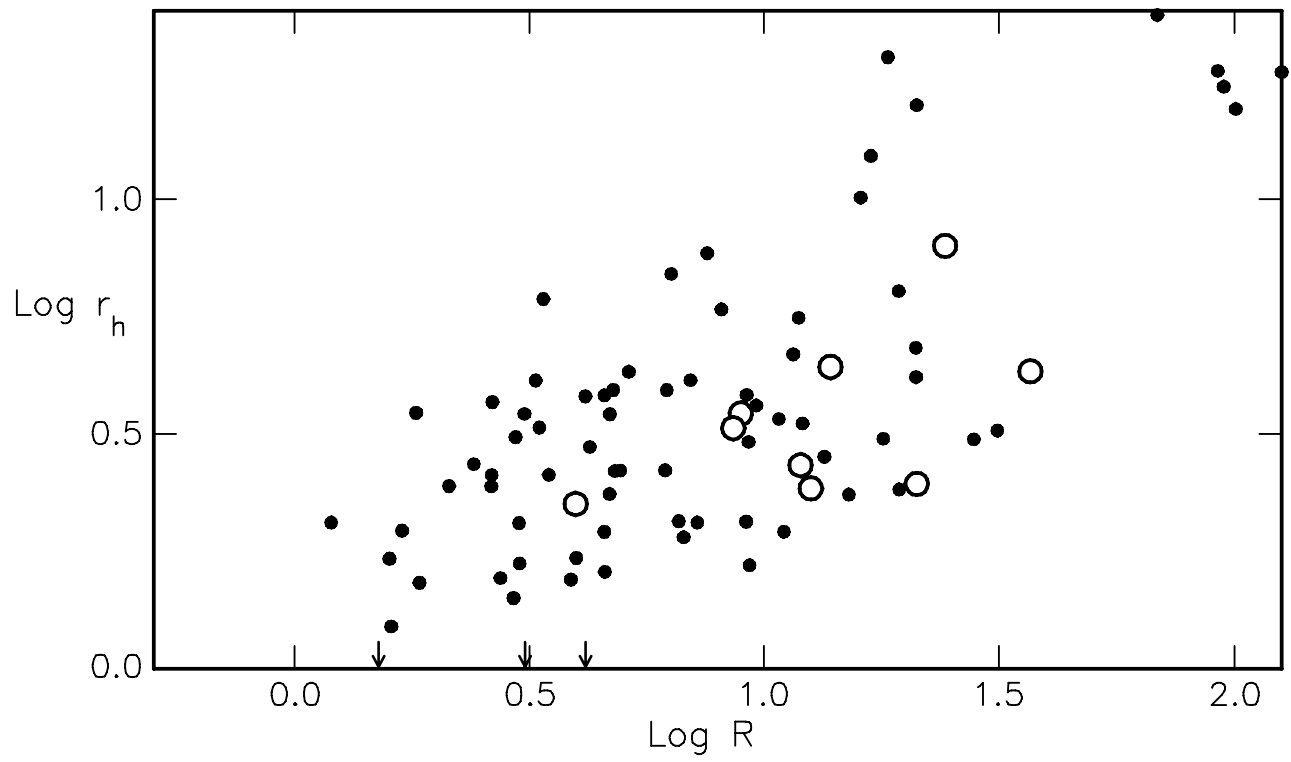